# Dynamic Metasurface Antennas: From Programmable Microwave Hardware to Next-Generation Wireless Systems

Abdul Jabbar* and William Whittow
Wolfson School of Mechanical, Electrical and Manufacturing Engineering, Loughborough University, UK
*a.jabbar@lboro.ac.uk

As wireless systems evolve toward 6G and beyond, antenna apertures are increasingly expected to become more programmable, multifunctional, and responsive to diverse communication, sensing, and imaging requirements. Dynamic metasurface antennas (DMAs) offer a promising hardware pathway toward this vision by integrating electronically addressable metamaterial elements within microwave and millimeter-wave (mmWave) radiating apertures. Unlike conventional phased arrays relying on dedicated phase-shifting networks, DMAs manipulate guided waves in microwave structures through electronically controlled resonant elements. In this way, DMAs offer phase-shifter-less beam-steering, multibeam radiation, and programmable wavefront control. This article explores state-of-the-art DMA technology from a microwave-engineering perspective, spanning guided-wave excitation, reconfigurable resonant meta-elements, microwave tuning technologies, RF-digital integration, and FPGA-controlled beamforming. Closely related electronically steered metasurface antenna architectures are also discussed within a unified microwave-hardware framework. Recent advances illustrating how guided-wave DMA hardware and high-speed electronics facilitate programmable communication, sensing, imaging, and space-time functionalities are reviewed. Key microwave design challenges toward wideband, scalable, and highly integrated DMA apertures are highlighted. Finally, emerging opportunities in satellite communications, near-field connectivity, fixed wireless access, wave-domain computing, and integrated wireless functionalities are explored as future directions for DMA technology. This article provides a unified microwave engineering perspective on the design, practical implementation, and emerging capabilities of DMAs as programmable front-ends for next-generation wireless systems.

## Introduction

Metamaterials and their two-dimensional (2D) planar counterparts, metasurfaces, have revolutionized the landscape of electromagnetic (EM) wave control over the last two decades [1]–[5]. Their significance stems from the unprecedented ability to tailor the amplitude, phase, polarization, and propagation characteristics of an incident or guided wave through the collective response of carefully designed subwavelength unit cells, offering functionalities that are difficult to realize using naturally occurring materials alone [6], [7]. Antenna engineering has been among the most fertile grounds for this idea. The pivotal step has been the move from static metasurface apertures to programmable ones whose radiation characteristics can be reconfigured electronically in real time. Among these programmable metasurfaces, the Dynamic Metasurface Antenna (DMA) has emerged as a leading embodiment, shifting electronic beam-steering from conventional phase-shifter-based architecture toward a metasurface-steered paradigm, as depicted in Figure 1. DMAs offer a hardware-efficient path to electronically reconfigurable microwave and millimeter-wave (mmWave) antennas, through embedded tunable metamaterial elements directly within the radiating structure itself. This integration is what sets them apart from conventional phased arrays, as shown in Figure 2. Rather than steering a beam with dedicated phase-shifting integrated circuits (ICs) and complex corporate feed networks, a DMA can eliminate the need for conventional element-level phase shifters while reducing feed-network complexity [8].

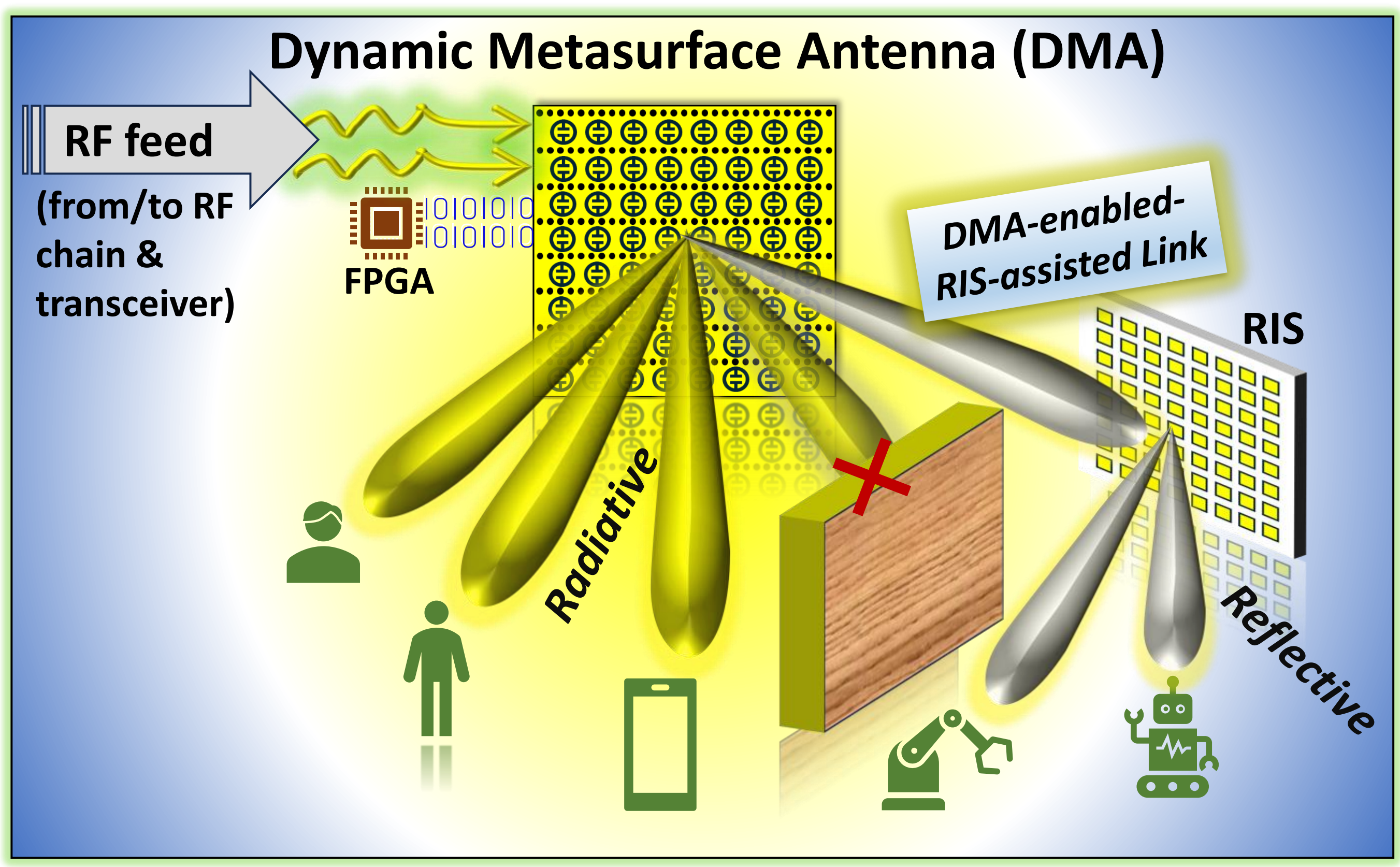

Figure 1. Conceptual illustration of phase-shifter-less electronic beam-steering enabled by a DMA, highlighting the microwave-hardware foundation of programmable beamforming and a prospective DMA-enabled RIS-assisted wireless scenario.

Legacy phased arrays have long served as the dominant beamforming solution for such systems. However, their dependence on dedicated phase shifters and complex feeding networks introduces substantial challenges in terms of power consumption, hardware cost, scalability, and implementation complexity [9]. As antenna apertures continue to grow in size and functionality, simply scaling the existing phased-array architectures may no longer represent a sustainable path toward future wireless networks. The emerging 6G vision suggests a fundamentally different perspective in which wireless networks become capable of actively shaping and reconfiguring their EM response in real time. In such systems, adaptability is no longer confined to radio-frequency (RF) signal processing algorithms but extends directly to the EM aperture itself.

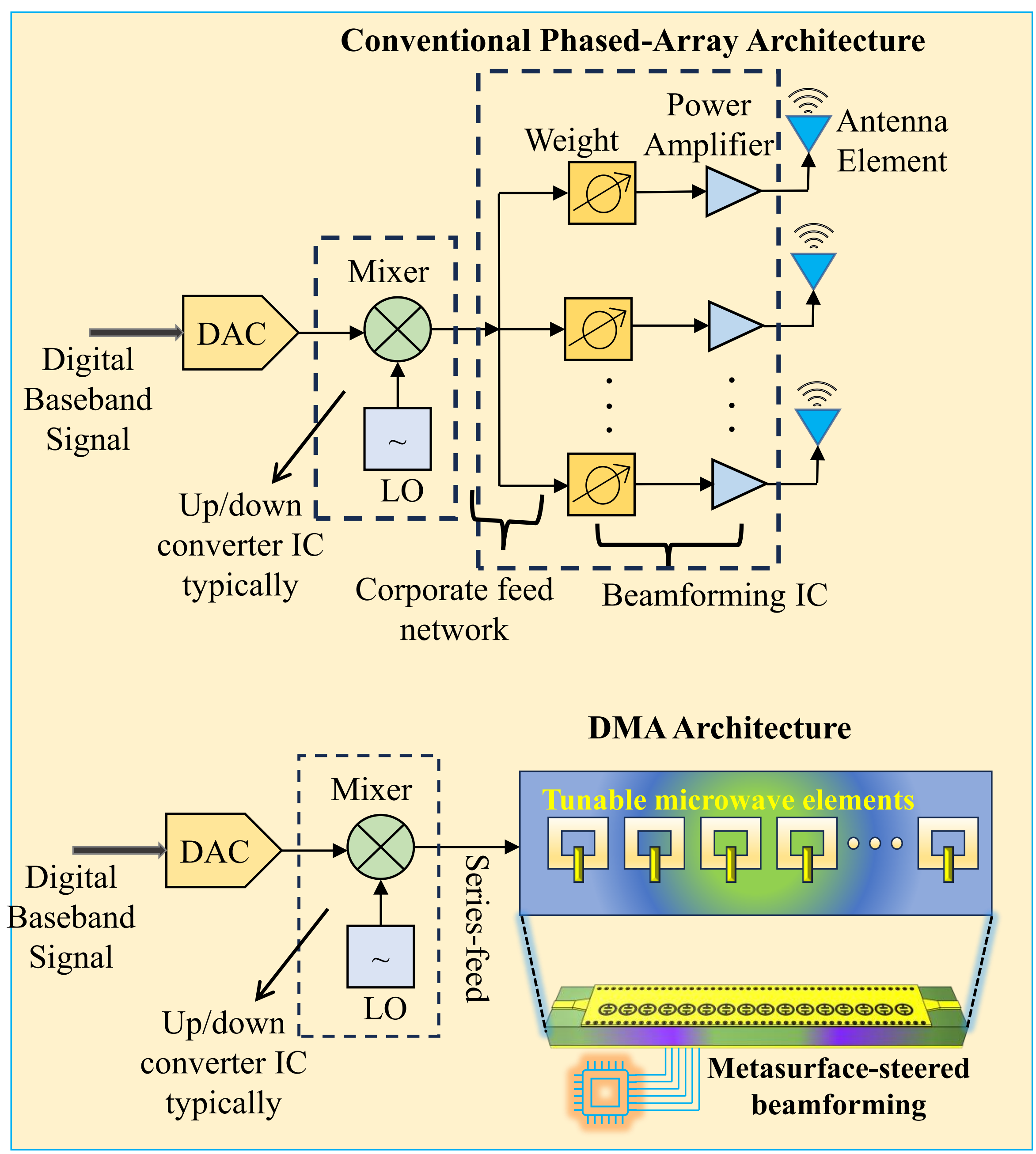

Figure 2. Illustration of the comparison between conventional phased-array (power-hungry phase-shifting network) and DMA (with software-controlled reconfigurable meta-elements and phase-shifter-less beamforming) radio architectures, (LO: local oscillator, DAC: digital-to-analog converter).

From a microwave-engineering perspective, a DMA is more than a reconfigurable antenna. It is a distributed microwave circuit in which a guided RF mode interacts with an array of electronically tunable resonant loads. Its performance therefore depends jointly on transmission-line propagation, resonator coupling, semiconductor-device impedance, RF/DC isolation, conductor and dielectric losses, switching electronics, and digital control. Understanding these interactions is essential for translating programmable metasurface concepts into practical RF front-ends. Table 1 compares the microwave hardware architectures and key implementation characteristics of conventional phased arrays and DMAs.

**Table 1. Hardware comparison of DMA and conventional phased-array architectures.**

| Attribute | Conventional phased array | DMA |
|---|---|---|
| RF signal distribution | Corporate/series feed | SIW-fed guided-wave feed |
| Element control | Phase/amplitude IC | Resonant impedance/loading |
| Phase shifters | Typically required | Not required at each element |
| RF chains | Architecture dependent | Potentially reduced |
| Beam Control | Beamformer ICs | Bias network + FPGA/controller |
| Primary loss sources | RF feed network, power amplifier, phase-shifter, interconnect | Guided-wave, resonator structure, energy loss to peripheral elements, tuning device |
| Bandwidth | Often comparatively broad band depending on architecture | Often resonance/dispersion limited, narrowband |

## Programmable Guided-Wave Metasurface Architectures: A Common Microwave Hardware Viewpoint

It is instructive to mention here that much of the apparent diversity in reconfigurable metasurface antennas stems from differences in terminology and theoretical formulation rather than fundamentally different microwave hardware. For instance, Reconfigurable Holographic Surfaces (RHSs), developed largely within the wireless-communications community, are commonly described through the language of holographic beamforming [10]–[13]. Like DMA, a specific class of RHS can have an edge-fed guided reference wave that propagates across the aperture and is locally modulated according to its relationship with a desired object wave, to synthesize the desired radiated wavefront.

From a microwave-hardware perspective, this architecture closely matches that of a DMA in the way that both employ edge-fed, leaky-wave-like apertures directly connected to an RF transceiver and populated by electronically tunable subwavelength radiators that progressively couple energy from the guided mode into free space [10], [12]–[14]. The connection becomes particularly evident in implementations based on tunable CELC resonators, whose interaction with a guided reference wave forms the physical basis of DMAs and RHSs. Hence, from a microwave-hardware perspective, an edge-fed RHS employing a guided reference wave and electronically tunable radiating elements can essentially be regarded as a realization of a DMA. Nevertheless, generic RHS architectures need not be restricted to edge-fed excitation. The guided reference wave can also be launched from an arbitrary location (other than edge) within the metasurface aperture and propagate as a guided wave, progressively coupling energy into free space through the tunable elements in a leaky-wave-like radiation mechanism [15].

Other DMA-like architectures appearing under different terminologies, including holographic metasurface antennas [16], and universal metasurface antennas [17], similarly exploit electronically controlled metasurface apertures in which a guided reference wave is spatially (and in emerging implementations, spatiotemporally) modulated to synthesize desired radiation patterns. Recognizing this shared EM lineage provides a useful bridge between research communities, allowing advances in resonator design, tuning technologies, biasing networks, aperture synthesis, and control algorithms to be viewed within a common framework. Throughout this article, we therefore consider these closely related edge-fed programmable metasurface architectures as a broader family, using the DMA as the representative hardware platform.

**Microwave Anatomy of a DMA**

A DMA can be viewed as a specialized class of leaky-wave antenna (LWA) structure in which EM energy is serially coupled from a guided wave into free space. However, unlike conventional LWAs that rely on fixed radiating slots inside a substrate-integrated waveguide (SIW) structure with limited or no electronic tunability, a DMA often leverages a complementary electric-inductive-capacitive (CELC) metamaterial element inside the SIW slot to achieve electronic reconfigurability in a software-programmable manner [18]–[20], as shown in Figure 3(a). The reconfigurability can be realized through several tuning mechanisms, including positive-intrinsic-negative (PIN) diodes for discrete, typically binary control, and varactor diodes for continuous tunable responses. Emerging alternatives, including liquid crystals and phase-change materials, also provide tunability, although their integration into practical DMA hardware remains comparatively less mature.

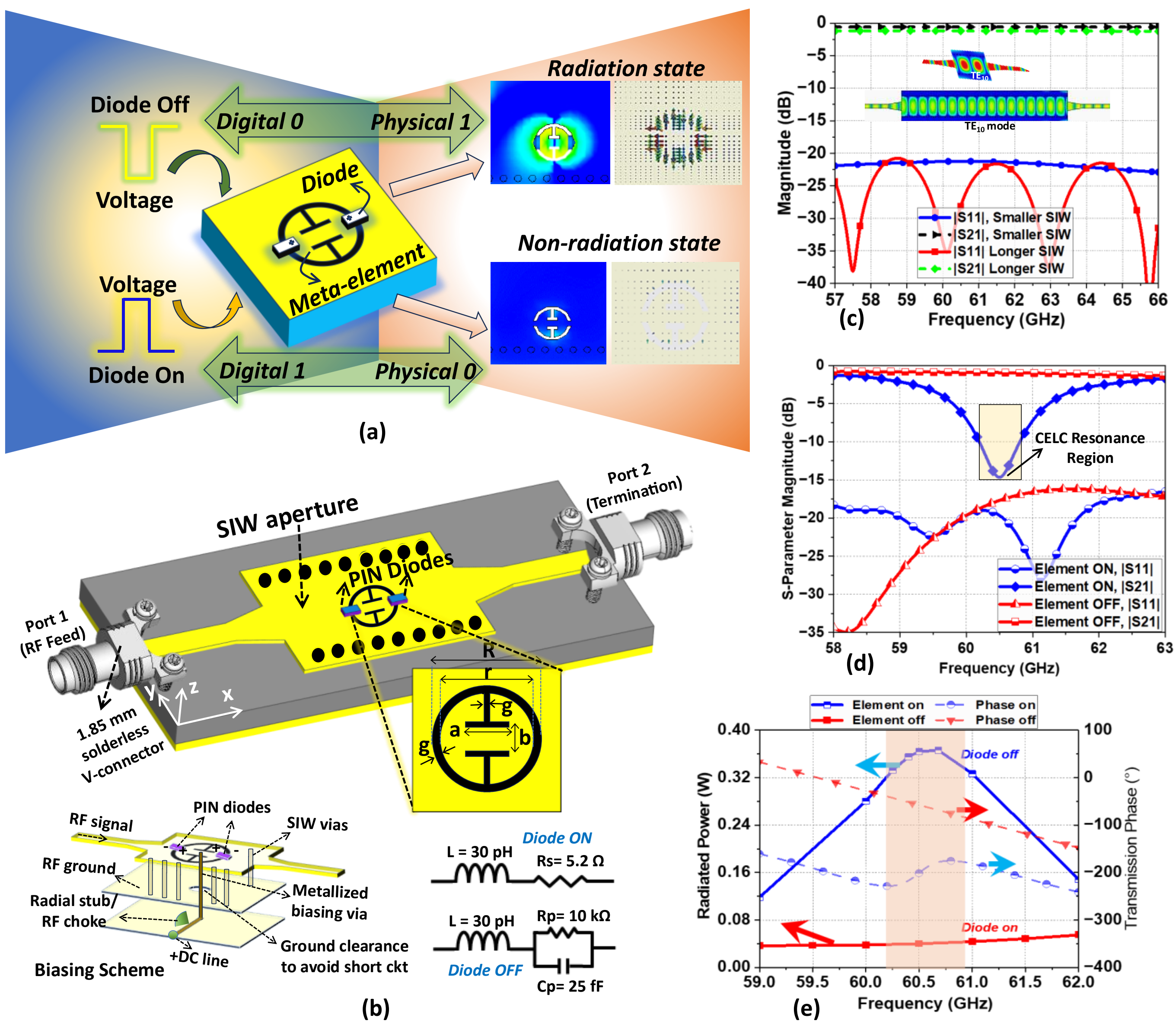

Figure 3. Microwave operation of a tunable DMA element. (a) PIN-diode-controlled CELC states and corresponding radiating/non-radiating behavior. (b) Schematic design of an SIW-fed DMA element with an equivalent microwave circuit model of the PIN diode in ON and OFF states. $R = 1.1$, $r = 0.9$, $g = 0.1$, $a = 0.3$, $b = 0.2$). Dimensions are in mm. (c) Simulated scattering parameters of the host SIW. (d) S21 and S11 response of the DMA element in radiating and non-radiating states. (e) Radiated power and transmission-phase responses of the CELC element in its ON and OFF radiation states [20].

### Designing the Microwave Guided-Wave Structure

The first step is to design a low-loss SIW structure to act as the energy-guiding platform in the desired band of interest. The design objective is to achieve low transmission loss (high $S_{21}$) with minimal reflection (low $S_{11}$) to maximize the RF power available for controlled radiation. As a representative example, Figure 3(c) shows the scattering parameters of a low-loss SIW designed around the 60 GHz mmWave band, providing the foundation for subsequent mmWave DMA operation [20]. The SIW aperture was designed using an RO3003 substrate with a dielectric constant of 3, a thickness of 0.25 mm, and a conductor thickness of 17.5 microns. It is important to carefully design a microstrip-to-SIW transition for better impedance matching.

Full-wave simulations of the SIW structure show operation in the dominant $TE_{10}$ mode. The magnitude of the reflection coefficient remains below -20 dB across a wide range (57-66 GHz), while the transmission coefficient remains above -0.25 dB. This confirms the use of a low-loss SIW structure to host resonant meta-elements for dynamic functionalities.

**Designing the Reconfigurable Meta-Element**

To understand the design philosophy of the DMA element, consider Figure 3(a). A CELC resonator is primarily a magnetically coupled resonant structure that can interact strongly with the appropriate in-plane magnetic-field component of the SIW mode. Such a reconfigurable meta-element serves as the fundamental building block that facilitates EM tunability. In the example shown here, the CELC element integrates with the SIW aperture and provides digital control via switching PIN diode states. By switching the state of the PIN diode, the CELC dynamically regulates the coupling of guided energy into free space, allowing adaptive radiation control. While its operation can be understood intuitively, achieving the desired radiation characteristics typically requires full-wave EM optimization, as multiple geometric and EM parameters jointly influence the overall DMA performance (Figure 3(b)). The capacitive gap between the CELC element and the SIW aperture provides an easy mechanism to incorporate a PIN diode (or varactor diode) to achieve reconfigurable EM characteristics.

When the PIN diode is forward-biased (PIN switched on), the CELC element is short-circuited with the SIW aperture and therefore its resonant radiation is strongly suppressed. This leads to the non-radiating state (element-OFF), providing a high transmission ($S_{21}$) and a lower reflection ($S_{11}$) coefficient, similar to the response of the unloaded SIW. Conversely, when the PIN diode is switched off (reverse-biased), the CELC element is isolated from the SIW aperture and effectively couples to the in-plane SIW wave. Consequently, it resonates near the designed operating frequency, and couples guided energy into free-space radiation. This corresponds to the element-ON state. Notably, in this CELC implementation, the radiating state occurs when the PIN diode is OFF. Hence, it consumes much less power than the phased-array antenna architecture.

The CELC can be represented by a resonant RLC branch coupled to the SIW transmission line. With the PIN diode OFF, the CELC resonance is restored and the resonant branch strongly loads the guided mode near the resonance frequency. It extracts energy from the SIW and couples it into free-space radiation, which manifests as a pronounced dip in $S_{21}$ (Figure 3(d)). When the PIN diode is ON, its low impedance shorts the CELC gap and suppresses the resonance, significantly reducing the loading of the SIW and restoring high through transmission.

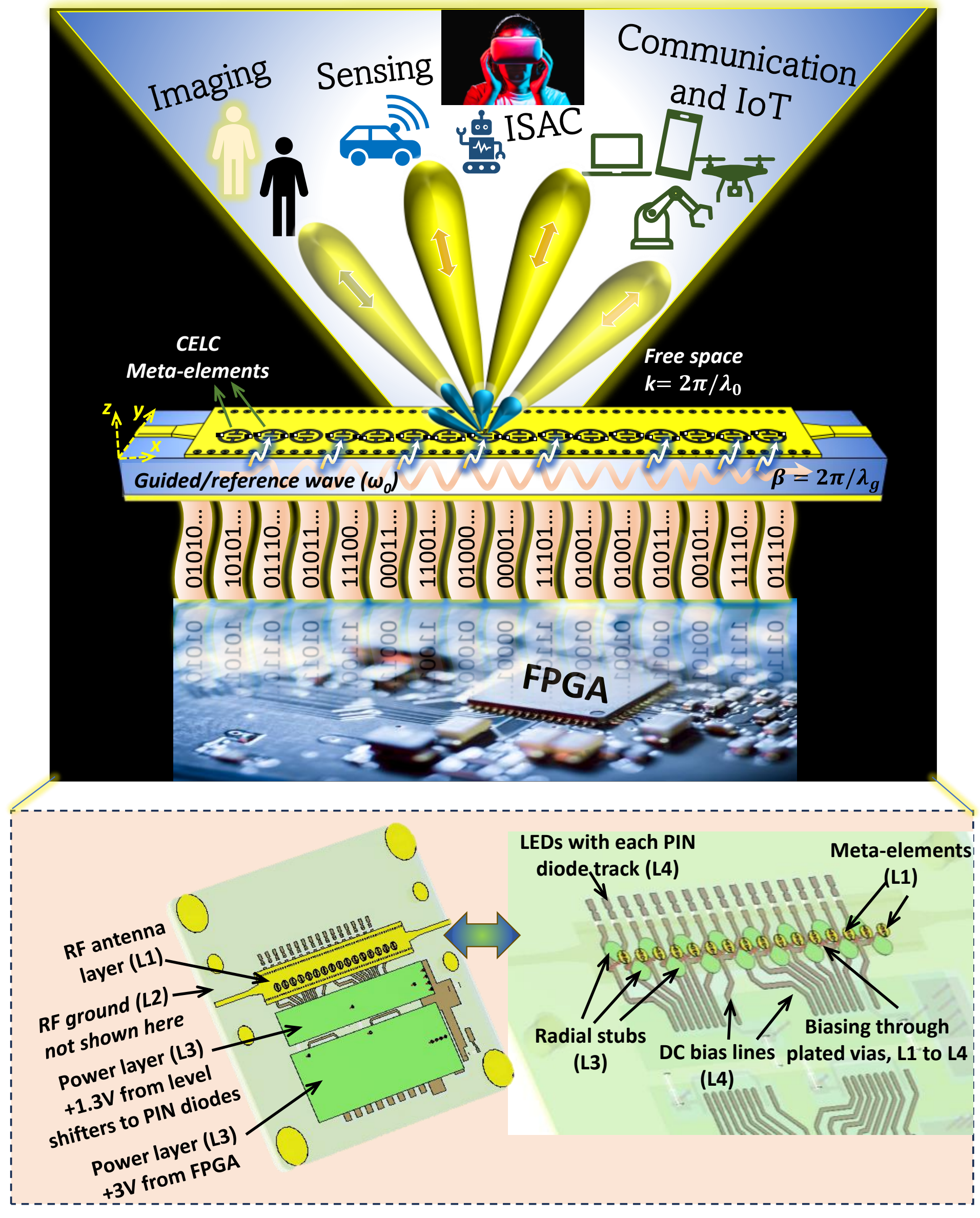

Figure 4. Conceptual depiction of a DMA with multilayer PCB stack-up, showing the integrated RF aperture, RF/DC isolation and biasing network, and electronic control interface [20].

## RF Power Flow and Radiation Profile

The RF power profile of the DMA is shown in Figure 3(e). For 0.5 W of input power, and 0.45 W of accepted power, a peak radiated power of 0.36 W is achieved at 60.5 GHz. The demonstrated radiation efficiency of DMA is approximately 80%. On the other hand, the element-OFF state shows radiation below 0.06 W for the same accepted power, corresponding to a radiation efficiency below 13.3%. Moreover, the transmission phase exhibits dispersive Lorentzian resonance around 60.5 GHz in the radiating state (confirming the resonance when the element is on), whereas the phase is linear in the non-radiating or element-OFF state.

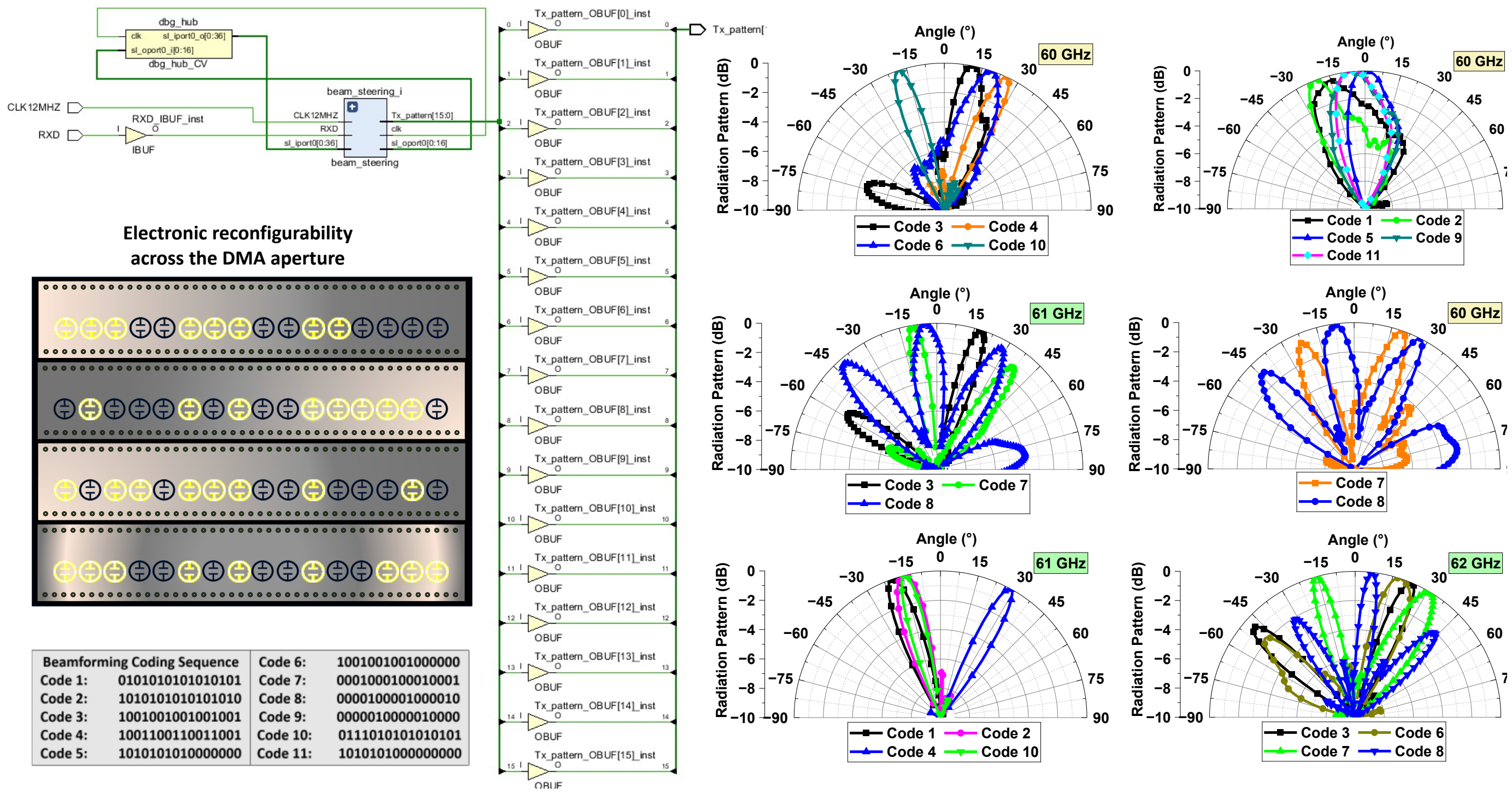

Figure 5. Demonstration of microwave aperture states of a DMA, along with a schematic diagram of the parallelized FPGA control scheme [21]. Different binary coding sequences dynamically reconfigure the metasurface aperture to generate distinct single- and multibeam radiation patterns across 60-62 GHz.

A potential hardware advantage of DMAs is the low DC power required for aperture-state control compared with the architectures employing large numbers of active phase-shifting ICs, as shown above in Figure 2. In the binary CELC implementation of DMAs, radiation occurs when the PIN diode is reverse biased, requiring negligible steady-state bias current. However, overall system efficiency also depends on guided-wave loss, resonator coupling, switching-device loss, aperture utilization, and the associated RF electronics. Selective activation of the meta-elements additionally provides a direct mechanism for modifying the effective aperture distribution and, consequently, the resulting beam shape and beamwidth.

## RF/DC Bias Network and Multilayer Integration

Building upon the fundamental DMA design principles, larger linear and planar apertures can be realized by integrating multiple reconfigurable meta-elements along the guided-wave structure. A representative 16-element DMA is illustrated in Figure 4. Practical DMA realization requires careful co-design across antenna and microwave engineering, multilayer PCB layout, DC biasing, and digital control. The meta-elements occupy the RF layer and couple to the guided reference wave, while the tuning devices provide electronically reconfigurable states of the elements. Each PIN diode requires an appropriate DC bias path, routed through dedicated bias lines and isolated from the RF network using high-impedance structures such as radial stubs.

**From Digital Bits to Microwave Aperture States: FPGA-RF Integration**

The binary control sequences are generated through a field-programmable gate array (FPGA) and routed through the multilayer PCB to independently configure the PIN-diode states. Integrating an FPGA into this architecture facilitates closed-loop control of the metasurface tuning state, allowing real-time beam-steering, null steering, or radiation-pattern adaptation based on the application requirements. Such RF-digital co-design is central to practical DMA implementation, since beam-steering performance depends not only on the metasurface geometry but also on reliable biasing, RF isolation, routing, and high-speed digital control.

As demonstrated in Figure 5, different binary beamforming coding sequences can generate single-, dual-, and multi-beam radiation patterns toward distinct spatial directions within ±50° [20]. It is worth noting that the available coding space in DMAs is remarkably large. As an example, a 16-element DMA can generate 65,536 possible coding combinations ($2^{16}$ = 65,536), supporting a rich set of radiation patterns and wavefront configurations. In practice, however, only a subset of these states is typically useful, highlighting the need for intelligent beamforming strategies and efficient biasing schemes to identify and realize optimal aperture configurations.

The FPGA integrates a synchronization circuit that generates the required clock and control signals while receiving beam-steering commands from a graphical user interface. Operating with a 100-MHz clock, the controller enables aperture-state reconfiguration with a minimum timing resolution of 10 ns. Binary coding sequences are transferred to the FPGA through a universal asynchronous receiver-transmitter (UART) interface, together with commands defining the operating mode, including single- or multibeam operation, beam-switching, and the switching interval between successive codes. This programmable control architecture allows synchronized and rapid updating of the DMA element states, and provides a hardware platform for dynamic beamforming and potentially more advanced spatiotemporal coding schemes. The overall architecture of one such FPGA control logic implemented in VHDL using Xilinx Vivado is illustrated in Figure 5 [21].

**Application-Driven Microwave Design of State-of-the-Art DMAs**

**DMAs for Wireless Communication**

For wireless communication, a DMA must not only form the desired beam but also deliver RF power efficiently to establish a reliable link. A well-controlled, low-loss guided-wave structure, such as an SIW, is therefore attractive for achieving good impedance matching, radiation efficiency, gain, and beam quality. Sidelobe levels (SLLs) must also be carefully controlled during the design process, as unwanted radiation can increase interference and degrade the link margin. These requirements become more challenging in multibeam operation, where the available aperture power is distributed among multiple spatial directions. Thus, a communication-oriented DMA should be designed from a complete RF front-end perspective, with aperture performance evaluated in the context of the overall link budget. Several DMA hardware implementations for wireless communication have been reported across a wide frequency range, including approximately 10 GHz [22], 12 GHz [23], 12.5 GHz [24], 22-30 GHz [16], [17], [25]–[27], 60 GHz [20], and 105 GHz [28], as summarized in Table 2.

The rapid and parallelized FPGA-based control has supported space-time modulation of DMA apertures, extending their functionality beyond purely spatial beamforming [16], [17]. By dynamically switching the metasurface states in both space and time, frequency harmonics with independently controllable radiation characteristics can be generated [29], [30]. Furthermore, the temporal coding sequence can itself be made information-dependent, so that modulation and beamforming are implemented jointly at the programmable aperture rather than at the conventional RF-modulation level. This can potentially reduce reliance on additional RF chains and conventional RF modulation/demodulation hardware.

**DMAs for RF Sensing**

DMAs act as a programmable illuminator that dynamically probe targets or the surrounding environment using reconfigurable spatial field patterns [31], [32]. The emphasis often shifts toward adaptive illumination and measurement diversity. Electronically steered beams from a DMA can be leveraged for radar-like sensing to sequentially interrogate different directions, detect and track moving targets, or monitor multiple subjects across a scene [33]. Unlike communication-oriented DMAs, however, a highly directive beam with consistently low sidelobes is not always required. Rather, depending on the sensing modality, field diversity and sensitivity to target-induced changes may be more important. Different aperture states provide distinct EM observations that can be processed to estimate parameters such as range, position, motion, or physiological activity [32], [34]. Nevertheless, it is instructive to mention here that the signal processing can only process information captured by the RF front-end. No amount of post-processing can recover a target signature that was not adequately captured in the received signal in the first place. This makes the EM design of the DMA aperture as important as the downstream sensing algorithms.

**DMAs for Microwave Imaging**

DMA-like apertures have been extensively utilized for computational imaging [35]–[39]. In an imaging use-

| Table 2. Representative state-of-the-art microwave hardware implementations of DMAs. | | | |
|---|---|---|---|
| **Reference** | **Operating Frequency (GHz)** | **Microwave Component for Reconfiguration** | **Demonstrated Function** |
| [22] | 10-12 | Varactor diode | ✓ Communication |
| [12], [14] | 11-12 | PIN diode | ✓ Communication<br>✓ ISAC |
| [23], [24] | 12-14 | PIN diode | ✓ Communication |
| [17] | 23 | PIN diode | ✓ Communication<br>✓ Space-time holography |
| [25] | 26 | Varactor diode | ✓ Communication |
| [26] | 27 | PIN diode | ✓ Communication |
| [27] | 29 | Liquid crystal | ✓ Communication |
| [20], [21], [32] | 60 | PIN diode | ✓ Communication<br>✓ RF sensing |
| [28] | 105 | Liquid crystal | ✓ Communication |
| [31] | 19.5 | Varactor diode | ✓ RF sensing (object sensing, fall detection) |
| [33] | 20 | PIN diode | ✓ RF sensing (handwriting and motion detection) |
| [40] | 20 | PIN diode | ✓ Integrated sensing and imaging |
| [41] | 22.5 | PIN diode | ✓ Near-field focusing imaging |
| [37], [42] | 19-24 | PIN diode | ✓ Computational microwave imaging |
| [16] | 24.5 | PIN diode | ✓ Microwave imaging<br>✓ Space-time holography |

case, a DMA operates as a coded EM aperture that generates spatially diverse [42]–[46] or frequency-diverse [47]–[50] field patterns to illuminate an object. Each pattern illuminates the scene differently, and the corresponding scattered fields provide a measurement that encodes information about the objects within the field of view [37], [44]. For this purpose, an aperture-launched or cavity-excited wave can be particularly useful because the objective is often to generate a rich collection of spatially diverse radiation masks rather than a single highly directive beam with a desired SLL.

The metasurface states effectively act as EM measurement masks, and an image is reconstructed computationally from the ensemble of measured responses [51], [52]. Therefore, for imaging applications, DMA performance should not be assessed solely through conventional antenna metrics such as gain or SLL. Instead, the DMA aperture can be optimized for pattern diversity, spatial coverage, and the EM information required for computational reconstruction [40], [41].

## DMAs for Space-Time Microwave Holographic Imaging

Microwave holography is a wavefront-reconstruction principle in which the interference between a known reference wave and a desired object wave forms a holographic pattern that encodes the amplitude and phase information of the object wave. In a DMA, the guided wave propagating across the aperture acts as the reference wave, while the desired free-space beam represents the object wave. The resulting holographic pattern determines the required states of the metasurface elements, which modulate the guided wave to reconstruct the desired radiated wavefront. This process is commonly referred to as holographic beamforming [13].

It is instructive to mention here that conventional computational imaging can exploit spatial or frequency diversity from quasi-static aperture states to encode scene information. In contrast, space-time holography deliberately introduces time-varying impedance states and harmonic generation in DMAs. Space-time holographic beamforming generates controlled frequency harmonics that provide additional temporal and spectral degrees of freedom for harmonic beam-steering, field synthesis, beam shaping, and image reconstruction [17], [41]. This places additional microwave hardware requirements on switching speed, FPGA control bandwidth, timing synchronization, harmonic conversion efficiency, and spectral purity, which are central considerations for DMA hardware.

Although the same programmable microwave-hardware foundation can support multiple functionalities, its design priorities change considerably with the intended EM operation, as summarized in Table 3.

## DMA Hardware Case Studies

### Microwave Space-Time Holographic Imaging

From a microwave-engineering perspective, a space-time holographic metasurface antenna using PIN

| Table 3. Application-driven microwave design considerations for DMAs. | | |
|---|---|---|
| **DMA Function** | **Key Microwave Design Priorities and EM Objectives** | **Microwave Hardware Considerations** |
| Wireless Communication | • Low insertion loss<br>• Impedance bandwidth control<br>• High radiation and aperture efficiency<br>• Beamwidth and sidelobe control<br>• Efficient beamforming control circuitry<br>• Fast beam-switching control (for low-latency operation) | • Low-loss SIW design<br>• RF feed transitions<br>• Resonator coupling<br>• Tuning-device loss<br>• RF/DC isolation design<br>• Parallelized FPGA control scheme |
| RF Sensing | • Spatial/illumination diversity<br>• Low-correlation illumination patterns<br>• Phase-coherent operation<br>• Repeatable aperture states<br>• Target interrogation and weak signature capture<br>• Fast aperture reconfiguration | • Phase-stable RF chain<br>• Repeatable element states<br>• Tx/Rx synchronization<br>• Low phase noise<br>• RF/DC isolation |
| Computational Imaging | • Wide spatial coverage<br>• Low pattern correlation<br>• Spatial/frequency diversity<br>• Reconstruction fidelity | • Cavity/guided-wave dispersion<br>• Broadband RF transitions<br>• Pattern characterization<br>• Aperture-state calibration |
| Space-Time Holography | • Spatiotemporal wavefront control<br>• Harmonic efficiency and spectral purity<br>• Timing synchronization<br>• Modulation bandwidth<br>• Harmonic generation and control | • High-speed microwave tuning devices<br>• High-speed bias/control network<br>• FPGA/driver bandwidth<br>• Switching transients and parasitics<br>• Timing/jitter control |

diodes is shown in Figure 6(a) [16]. It provides an illustrative example of how the guided-wave network, active RF elements, and digital control electronics can be co-designed to realize programmable holographic imaging. A continuous-wave reference signal at 24.5 GHz is distributed across a 2D aperture using a meandering SIW, whose rounded bends are designed to suppress higher-order modes and minimize reflections (Figure 6(b and c)). Slot-type meta-atoms etched along the SIW extract the guided energy into free space, while two PIN diodes embedded in each meta-atom electronically switch its radiating state. The design includes 78 meta-elements which are independently controlled through an FPGA with a 1 kHz time-modulated signal to implement 3D binary space-time-coding matrices.

Importantly, the SIW performs both RF power distribution and guided-wave excitation, avoiding a separate 2D feed network and thereby reducing footprint and insertion loss. Space-time modulation then introduces the temporal degree of freedom in the way that the FPGA dynamically switches the meta-atoms according to synthesized space-time-coded sequences. It allows a single monochromatic RF excitation to reconstruct fields at different harmonic frequencies. The final imaging demonstration in Figure 6(d) shows that this microwave-digital architecture can reconstruct multifrequency 2D images and a 3D holographic object in the near field. This demonstrates how SIW microwave engineering, PIN-diode switching, high-speed digital control, and computational field synthesis can be integrated within a single programmable DMA-like microwave aperture.

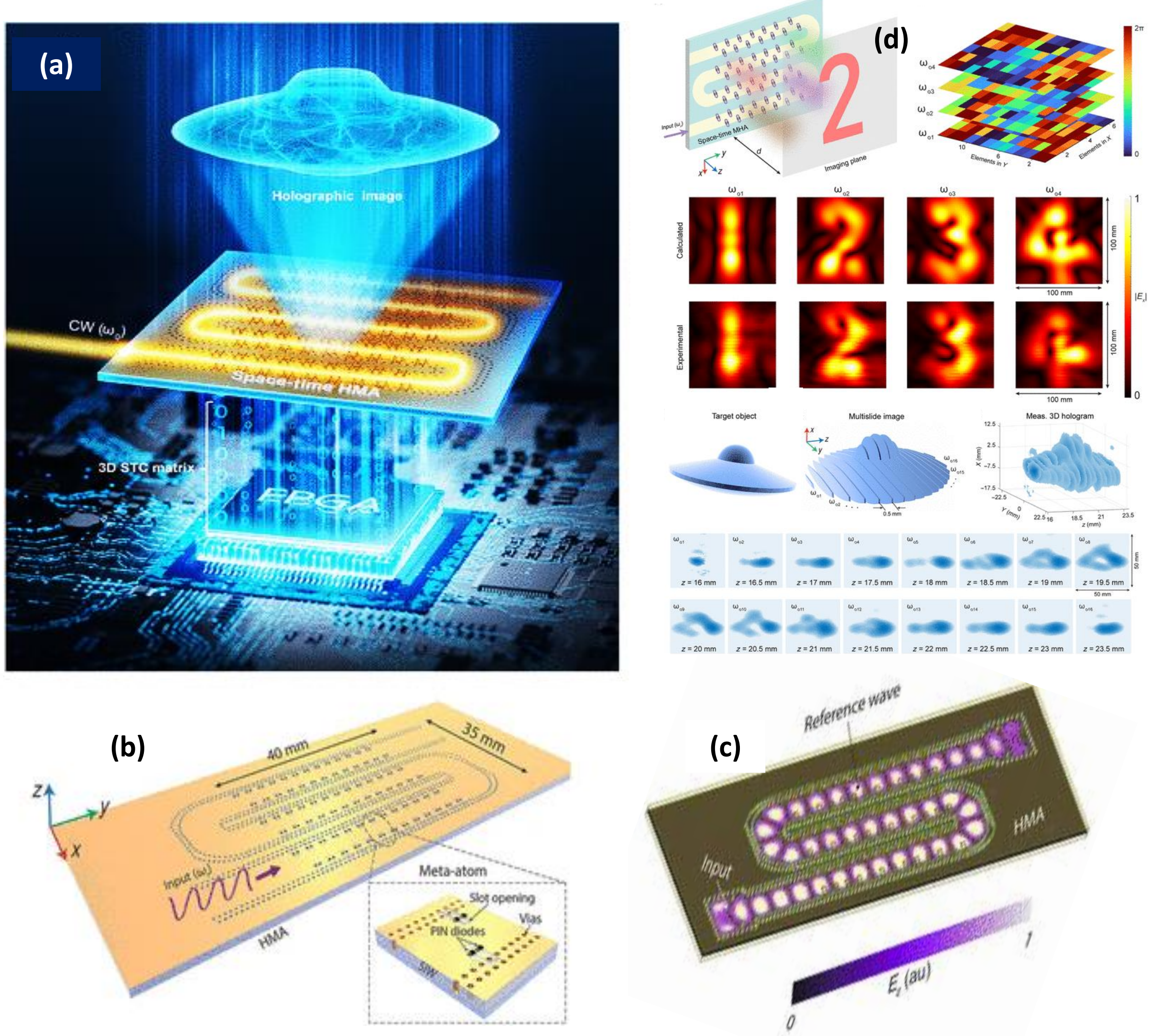

Figure 6. (a) Microwave holographic imaging with a DMA-like holographic metasurface antenna in which meta-atoms are modulated in both space and time from the integrated FPGA. (b) Meandered SIW-based design configuration. (c) Simulated electric field distribution of the reference wave inside the meandering SIW when all meta-atoms are in non-radiating states at 24.5 GHz. (d) Near-field holographic imaging with multi-sliced images and the experimentally reconstructed 3D holographic image [16] (CC BY-NC).

Note that the same programmable microwave DMA hardware can be configured differently according to the required electromagnetic function. Different aperture states and operating strategies can therefore be tailored to near-field focusing imaging [41], and other holographic integrated functionalities [17].

**DMA-Based Wireless Communication Testbed**

To demonstrate the practical feasibility of the DMA, a real-time 60 GHz communication testbed was developed using a DMA as the transmitter and commercially available mmWave front-end modules, as demonstrated in Figure 7 [21]. The DMA was configured through FPGA-based control to generate different radiation patterns and dynamically steer beams toward intended users. A high-definition quadrature phase-shift keying (QPSK)-modulated video stream was transmitted using a DVB-S2 communication framework, providing a realistic assessment of communication performance.

The DMA was configured with code 2 via the FPGA for the single-user scenario to generate a single beam toward -9° at 62 GHz to serve a single dedicated user. The distance between the Tx DMA and Rx horn antenna was set to 1 m. The real-time video transmission was received with a resolution of 1280 × 720 and an average transmission rate of 3.2 Mbps. Link quality was assessed directly from the SDRangel DATV demodulator, which reports the modulation error ratio (MER) and carrier-to-noise ratio (CNR) of the recovered QPSK stream in real time. Across repeated captures, the received signal exhibited an MER of approximately 12-13 dB and a CNR of approximately 6 dB, with the four QPSK clusters clearly resolved and the HD video decoded without interruption. Since MER is the digital-domain counterpart of the error-vector magnitude ($EVM\ (\%) = 100\times10^{-MER/20}$), it corresponds to an EVM of approximately 22-25%. Moreover, from the measured MER, an uncoded QPSK bit-error rate (BER) on the order of $10^{-5}$ can be estimated under an ideal Gaussian noise model using the standard QPSK relation, consistent with the artifact-free video reception.

To demonstrate multicast communication capability, the DMA was reconfigured to generate dual beams simultaneously using code 3, allowing two spatially separated receivers positioned at +23° and -49° to receive the same video stream concurrently. This capability highlights one of the key abilities of DMAs to dynamically reshape the effective aperture and redistribute EM energy without relying on conventional phase-shifter-based beamforming networks.

**Microwave Hardware Design Challenges**

Despite the promising capabilities of DMAs, several fundamental challenges remain unsolved that should be explored and addressed. In this section, we discuss the key open hardware design limitations of DMAs. We also highlight emerging research directions and applications of DMAs that may shape their role in 6G and beyond wireless networks.

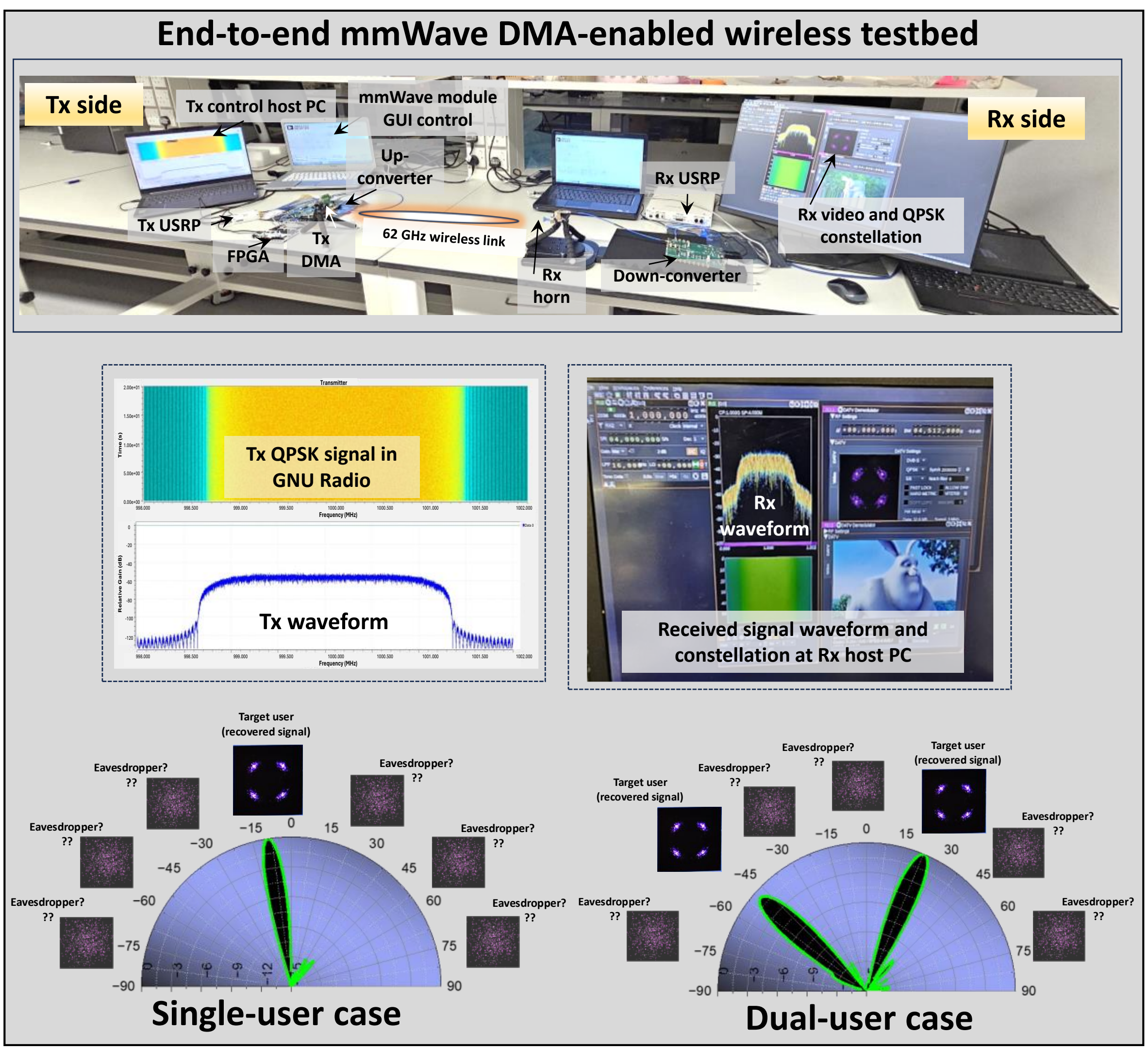

Figure 7. Experimental demonstration of programmable mmWave communication testbed using a DMA [21]. Electronic reconfiguration of the metasurface aperture facilitates real-time HD video transmission in both single-user and dual-receiver configurations through dynamically controlled single- and multibeam radiation.

***Wideband and Circularly Polarized DMAs:*** One of the major design challenges in DMAs is impedance and gain bandwidth. Most DMAs rely on resonant meta-elements, which naturally provide strong radiation control only over a limited frequency range. This becomes more critical for wideband 6G links, where a fixed coding state may not produce the same radiation response across the whole bandwidth. Future DMAs must therefore move beyond narrowband resonant operation toward wideband and dispersion-aware design techniques. Additionally, wideband DMA operation remains limited by the dispersive guided-wave response, which can induce frequency-dependent beam squint across large bandwidths and therefore requires further exploration and mitigation for wideband communication systems. Furthermore, circular polarization is particularly important for satellite communications, and other links where polarization mismatch can degrade the received signal. Novel techniques for realizing broadband, circularly polarized DMAs are therefore highly desirable but remain challenging.

***Radiation Efficiency and Dual-Feed Excitation:*** Enhancing radiation and aperture efficiency remains a key challenge for DMAs. Beyond mere beam-steering, practical communication systems require high gain, low sidelobe levels, and efficient aperture utilization across different coding states. Since only a subset of DMA meta-elements typically radiates for a given binary sequence, the effective aperture is often underutilized, reducing overall efficiency and gain. Future research should focus on the development of efficient DMA-specific design metrics that jointly optimize EM performance, communication quality, and beam control. In addition, a dual-feed architecture might offer a promising direction to improve aperture utilization, increase radiation efficiency, and enhance beam manipulation by exploiting bidirectional wave propagation and feed diversity.

***Scaling Toward THz Bands and Tunability Issues:*** At sub-THz and THz frequencies, the reconfigurable hardware design challenges for DMAs become increasingly severe. Although shorter wavelengths allow compact apertures accommodating many elements, the physical space available for tuning devices, bias lines, packaging, and thermal management becomes extremely limited. Conventional PIN diodes and varactor diodes may no longer provide the required performance above 100 GHz due to loss, parasitics, nonlinearity, and fabrication tolerances. Future THz DMAs therefore require new tuning technologies, such as MEMS, liquid crystals, graphene, phase-change materials, photonic control, or hybrid electronic-photonic biasing. In this view, THz DMAs cannot be addressed simply by scaling existing microwave designs, but rather they will require a fundamentally new hardware design strategy.

**Emerging Applications and Research Trends**

***DMA-enabled RIS-assisted Conformal Marriage:*** A reconfigurable intelligent surface (RIS) is a reflective microwave metasurface comprising electronically tunable elements that dynamically control the reflection characteristics of incident electromagnetic waves [6]. From a microwave-hardware perspective, RISs and DMAs follow fundamentally different design philosophies. The reflection-mode RIS elements are typically engineered for a high reflection magnitude (high $S_{11}$) with controllable phase across the operating band, whereas a DMA requires low input-port reflection coefficient (low $S_{11}$) to maintain impedance matching and efficiently launch RF power into the guided-wave aperture. Moreover, unlike RIS elements illuminated by an externally incident free-space wave, DMA meta-elements must be electromagnetically coupled to an in-plane guided reference wave and therefore require a carefully designed host structure, such as an SIW, rather than a microstrip patch element configuration. These contrasting requirements make the integration of RISs and DMAs a microwave co-design problem in the same operating band involving reflection control on the RIS side and impedance matching, guided-wave propagation, resonant coupling, and controlled radiation on the DMA side.

Although both of these antenna technologies are separate active areas of research, their integration creates a promising "DMA-enabled RIS-assisted" architecture, as depicted in Figure 1. This extends EM programmability from the transmitting aperture into the wireless environment. Such DMA-RIS integration can introduces new microwave co-design opportunities for coverage extension, blockage mitigation, localization, and adaptive propagation control.

***DMA-Enabled MIMO Communication:*** Multiple-input multiple-output (MIMO) technology has unlocked the spatial domain at the physical layer (PHY) to transform network capacity and spectral efficiency [53]. Building upon this foundation, massive MIMO and gigantic MIMO have also emerged as defining technologies for beyond 5G wireless networks by leveraging large-scale beamforming antenna arrays to provide highly directional transmissions and spatial multiplexing gains [54]. DMA is a promising candidate to enable next-generation MIMO operation by exploiting the spatial degrees of freedom of a programmable metasurface aperture to generate multiple independently controlled beams [55]. Independently fed SIW-based channels can serve as distinct MIMO ports where the tunable meta-elements dynamically reshape each aperture pattern to improve spatial separation, beamforming gain, channel diversity, or multiuser coverage [8].

A theoretical study of a DMA-enabled holographic MIMO reported that holographic beamforming with DMA can increase the maximum beamforming gain by at least 50% compared with a conventional hybrid analog-digital MIMO array of the same size [15]. Moreover, theoretical and simulated studies have demonstrated that planar DMAs can achieve a higher sum rate than a conventional phased array of the same size, offering increased data capacity with potentially lower hardware complexity and cost [11]. This requires to design practical large DMA apertures to reduce the hardware complexity associated with fully digital massive-MIMO arrays [56]. However, it is important to note that a single-feed DMA with one RF chain producing multiple simultaneous beams is a case of multicast communication, as depicted in [21], and does not by itself constitute spatially multiplexed MIMO. Rather, independent data streams generally require equivalent independent RF/baseband channels or alternative mechanisms. Space-time modulated microwave DMAs offer one such alternative by using time-varying metasurface reconfigurations to encode information across both spatial and temporal domains [17].

***LEO Satellite Applications and NTN:*** Low size, weight, power, and cost (SWaP-C) DMAs represent a promising architecture for future LEO satellite communications and non-terrestrial network (NTN) links, where mass, DC power, hardware volume, and thermal budgets are tightly constrained. These constraints are typically more stringent than those encountered at terrestrial base stations. DMAs could provide an alternative to electronically steered phased arrays through phase-shifter-less beam-steering, reduced feed-network complexity, and adaptive aperture control. Their potential advantages include reduced RF complexity, lower DC power consumption for beam control, and flexible beam shaping for tracking moving satellite links [57], [58]. Thus, DMAs could potentially support agile uplink and downlink satellite communication in future LEO constellations, as illustrated in Figure 8(a). As NTNs become increasingly integrated with future wireless systems, DMAs could therefore evolve from predominantly terrestrial communication apertures toward a programmable-antenna platform spanning both terrestrial and spaceborne connectivity.

***Integrated Sensing and Communication:*** DMAs are increasingly being investigated not as pure communication antennas but as integrated apertures to support ISAC for 6G and beyond networks [59]. Although signal-processing frameworks for DMA/RHS-enabled base stations are becoming increasingly mature, practical hardware implementations and their systematic benchmarking against conventional phased arrays, particularly for ISAC, remain at an early stage of development. DMAs can dynamically reshape spatial fields, illuminate spatial regions, and generate diverse radiation beams that can be leveraged independently for communication and radar-like sensing (Figure 8(b)). In addition, FPGA-controlled space-time modulation introduces temporal degrees of freedom that can be exploited for harmonic beam control, information modulation, and potentially simultaneous communication and sensing functions [16], [17], [20], [29], [52]. This makes them attractive candidates for ISAC. While a wide range of reflective space-time-coded metasurfaces have been demonstrated in the literature [60]–[62], these surfaces are typically externally illuminated and cannot be directly interfaced with an RF transceiver as a transmitting antenna. In contrast, the hardware realization of space-time-coded DMAs, in which a space-time-modulated signal can be generated and radiated directly from a guided-wave-fed programmable aperture, remains in its early stages of development, particularly in the higher mmWave bands.

***Near-Field Connectivity, mmWave FWA, and Wireless Backhaul:*** Near-field communication with extremely large DMA arrays is another important research frontier. As base station apertures become electrically large at mmWave and sub-THz frequencies, users may occur in the radiating near field rather than the far field. In this regime, DMAs can focus energy at specific spatial points rather than merely steering beams in angular directions. This opens opportunities for user-specific energy focusing, spatial multiplexing, physical-layer security, and holographic communication [63]. However, it also creates new challenges in near-field DMA-based channel modeling, codebook design, calibration, and real-time optimization of large planar DMA apertures.

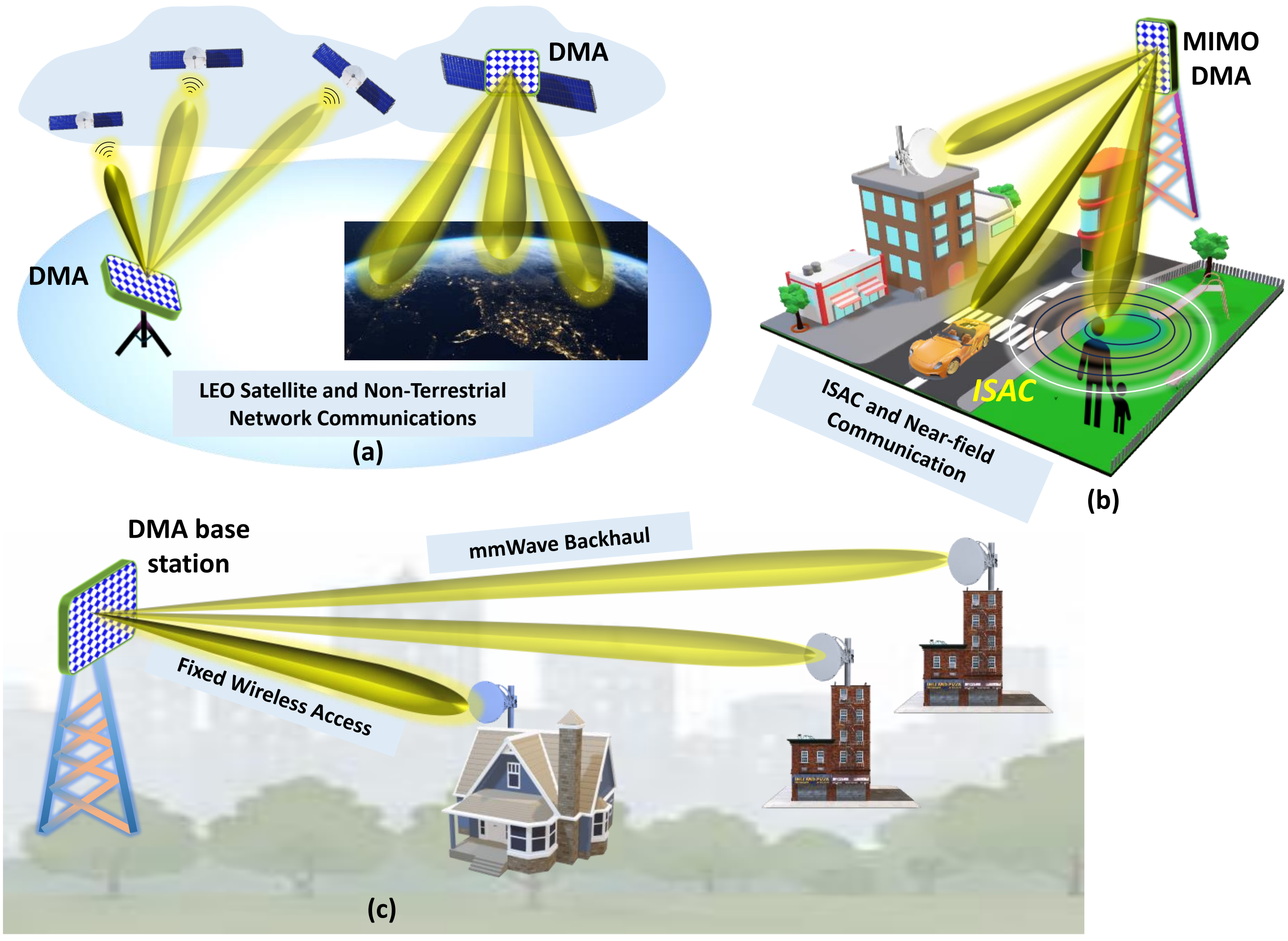

Figure 8. Prospective application landscape of DMAs in future wireless systems: (a) electronically steerable single- and multibeam links for LEO satellite and non-terrestrial networks; (b) programmable communication and sensing beams for ISAC and near-field connectivity; and (c) high-gain electronically steerable links for mmWave fixed wireless access and point-to-point wireless backhaul.

Fixed wireless access (FWA) and point-to-point mmWave wireless backhaul are other promising deployment scenarios for DMAs, particularly where fiber deployment is costly, difficult, or impractical. At mmWave frequencies, DMAs could serve as programmable base-station or terminal apertures, providing agile beam-steering and multibeam coverage with potentially lower RF hardware complexity than conventional phased arrays (Figure 8(c)). Future mmWave DMA architectures could support base stations and access points offering a hardware-efficient route toward high-capacity mmWave access and backhaul connectivity [64].

***Analog and Wave-Domain Computing:*** Analog and wave-domain computing represents another emerging opportunity for DMAs. By appropriately configuring the meta-elements, spatial transformations, filtering, combining, or wavefront operations could potentially be implemented before conventional digital processing [65], [66]. This approach is particularly attractive for large DMA-based MIMO, sensing, and computational imaging, where processing high-dimensional RF data can impose substantial hardware and computational overhead. Rather than digitizing and processing high-dimensional aperture data entirely in the digital domain, analog computing offloads selected operations to the physics of the programmable metasurface itself, effectively trading EM complexity for digital simplicity [67]. Space-time-controlled DMAs could further extend this concept by introducing programmable temporal and spectral transformations. However, practical DMA-based analog computing remains in its early stages, with challenges in accuracy, calibration, bandwidth, reconfigurability, and microwave RF-digital co-design.

## Conclusion

In this article, a microwave-hardware perspective on DMAs as programmable apertures for future wireless systems is presented. Beginning with the guided-wave host structure and reconfigurable meta-elements, the complete hardware chain, from resonant coupling and electronic biasing to RF–digital integration, FPGA-controlled beamforming, and end-to-end mmWave communication, is described. Application-driven microwave design priorities across wireless communication, RF sensing, and computational imaging are reviewed. Moreover, emerging opportunities in space-time modulation, ISAC, near-field operation, LEO satellite links, fixed wireless access, and wave-domain computing are highlighted. Although DMA architecture can reduce reliance on conventional phase-shifting networks, their broader adoption will depend on overcoming microwave-hardware design challenges associated with bandwidth, efficiency, tuning-device performance, and scalable RF-digital integration. Addressing these challenges will be critical to translating DMAs from promising laboratory prototypes into practical programmable microwave front-ends for next-generation wireless systems.